\documentclass[letterpaper,11pt]{article}

\usepackage{amsmath,amsfonts}
\usepackage{graphicx}
\usepackage{textcomp}
\usepackage{bmpsize}
\usepackage{xcolor}
\usepackage{lipsum}

\usepackage[ruled,vlined,linesnumbered,noend]{algorithm2e}
\usepackage{booktabs}
\usepackage{threeparttable}
\usepackage{enumerate}
\usepackage{graphicx} 
\usepackage{caption}
\usepackage{subcaption}
\usepackage{multirow}

\usepackage{tikz}

\def\BibTeX{{\rm B\kern-.05em{\sc i\kern-.025em b}\kern-.08em
    T\kern-.1667em\lower.7ex\hbox{E}\kern-.125emX}}

\begin{document}

\title{DeTrust-FL: Privacy-Preserving Federated Learning in Decentralized Trust Setting}

\author{
    Runhua Xu, Nathalie Baracaldo, Yi Zhou, Ali Anwar, Swanand Kadhe, and Heiko Ludwig
    \thanks{IBM Research Almaden, San Jose, CA, USA,
    {\tt \{runhua, yi.zhou, ali.anwar2, Swanand.Kadhe\}@ibm.com, \{baracald, hludwig\}@us.ibm.com}}
    }

\date{}
\maketitle

\begin{abstract}
Federated learning has emerged as a privacy-preserving machine learning approach where multiple parties can train a single model without sharing their raw training data.
Federated learning typically requires the utilization of multi-party computation techniques to provide strong privacy guarantees by ensuring that an untrusted or curious aggregator cannot obtain \textit{isolated} replies from parties involved in the training process, thereby preventing potential inference attacks.
Until recently, it was thought that some of these secure aggregation techniques were sufficient to fully protect against inference attacks coming from a curious aggregator.
However, recent research has demonstrated that a curious aggregator can successfully launch a \textit{disaggregation attack} to learn information about model updates of a target party.
This paper presents \textit{DeTrust-FL}, an \textit{efficient privacy-preserving federated learning framework} for addressing the lack of transparency that enables isolation attacks, such as disaggregation attacks, during secure aggregation by assuring that parties’ model updates are included in the aggregated model in a private and secure manner.
\textit{DeTrust-FL} proposes a decentralized trust consensus mechanism and incorporates a recently proposed decentralized functional encryption scheme in which all parties agree on a \textit{participation matrix} before collaboratively generating decryption key fragments, thereby gaining control and trust over the secure aggregation process in a decentralized setting.
Our experimental evaluation demonstrates that DeTrust-FL outperforms state-of-the-art FE-based secure multi-party aggregation solutions in terms of training time and reduces the volume of data transferred. In contrast to existing approaches, this is achieved without creating any trust dependency on external trusted entities.
\end{abstract}


\section{Introduction}

Training machine learning (ML) models that generalize well and have good predictive accuracy requires good quality training data. 
Traditionally, to train ML models, data is collected in a central place by the entities that run the training process.
This data oftentimes includes private information, whose use and transmission have been restricted by regulations including the Health Insurance Portability and Accountability Act (HIPPA), the European General Data Protection Regulation (GDPR), and California Consumer Privacy Act (CCPA).
To address privacy concerns and satisfy regulatory compliance, federated learning (FL) \cite{konevcny2016federated} has emerged as a promising learning paradigm, where multiple parties can collaboratively train a single machine learning model without sharing any of its training data.
In other words, while traditional machine learning systems require the transmission of all training data to a single place, FL ensures that each data owner or party can maintain possession of their own data.
For this reason, FL represents a substantial privacy improvement.

In FL, an entity called \textit{aggregator} is introduced to orchestrate the learning process among multiple parties who own private data. 
To train a model in FL, the aggregator sends an initial model to all parties, which then continue training it locally using their private data. Then, each party sends to the aggregator model parameters, e.g., gradients, which are known as \textit{model updates} \cite{ludwig2020ibm}.
The aggregator then combines or fuses all model updates received creating a single shared ML model.
These steps are repeated multiple \textit{rounds}, and in each of them, the aggregator shares with the parties a new model and parties train and send back model updates. Private data is never transmitted from the party; only model updates are shared with the aggregator.

In FL all training data remains with the party who owns it, providing basic privacy.
However, this simple FL setup still leaves room for \textit{inference attacks} that enable adversaries to obtain
information about the training data used by each party \cite{nasr2019comprehensive, shokri2017membership}.
The risk of private information leakage may be intolerable in some scenarios, and hence, multiple defenses
to prevent these threats have been proposed including differential privacy (DP) and secure multi-party computation (SMC).

SMC approaches protect privacy of the input through cryptographic techniques
to ensure that a curious or mistrusted aggregator cannot see individual model updates.
Popular SMC approaches include (partial) homomorphic encryption \cite{liu2019secure, zhang2020batchcrypt}, threshold Paillier \cite{truex2019hybrid}, functional encryption \cite{xu2019hybridalpha}, and pairwise masking protocols \cite{bonawitz2017practical}.
In contrast,  to mitigate inference attacks that use the final model or model updates, 
DP injects a carefully tuned amount of noise into model updates using differentially private mechanisms.
While DP provides strong privacy guarantees, approaches relying on DP are notorious for producing noisy models with lower accuracy.
Other solutions to prevent data leakage cleverly combine DP and SMC techniques to provide strong differential private guarantees while still enabling good model performance \cite{truex2019hybrid}.

Unfortunately, existing SMC techniques are not sufficient to prevent inference threats for the following reasons.
First, the lack transparency in the aggregation process results in an absence of
guarantee that all of the received model updates will be included in a certain round of the FL process.
Hence, a malicious aggregator can avoid including a target model update(s) in the secure
aggregation process leading to multiple problems.
Besides enabling inference attacks, a model produced by ignoring certain model updates may not generalize well for all participants or lead to unfairness in its predictions.
In an \textit{isolation attack} \cite{chen2011secure,liu2019boosting},
a malicious aggregator may try to segregate certain replies
by ignoring some of the received model updates to later on infer private information from a target party, thus violating its privacy. These attacks completely overthrow the purpose of SMC. 
Additionally, secure aggregation techniques are vulnerable to \textit{disaggregation attacks} \cite{lam2021gradient, so2021securing}, where a malicious aggregator can take advantage of the multi-round nature of federated learning to construct a system of equations to ultimately solve for the model updates of a target party. 
In summary, using existing crypto-based secure aggregation systems,
parties cannot be sure that all the received model updates have been aggregated as expected and are vulnerable to potential disaggregation leading to inference attacks.

To address the previously described limitations, we propose \textit{DeTrust-FL}, an efficient, scalable, and secure aggregation based privacy-preserving \underline{f}ederated \underline{l}earning \underline{de}centralized \underline{trust} approach, that solves the dilemma of secure multi-party aggregation efficiency and centralized trust.
\textit{DeTrust-FL} still follows the direction of advanced cryptography based approach to maximize simplicity and efficiency in terms of communication and computation, and is built on a recently proposed decentralized functional encryption (FE) \cite{abdalla2019decentralizing, chotard2018decentralized} design in which each party generates the functional decryption key collaboratively.
The aggregator can only obtain the aggregated result from the received encrypted parties' model updates using the functional decryption key.
As a result, \textit{DeTrust-FL} overcomes the limitation of previous state-of-the-art cryptography-based secure aggregation solutions.


Our key contributions are summarized as follows:

\begin{itemize}
    \item We propose \textit{DeTrust-FL}, an efficient and privacy-preserving federated learning (PPFL) in a decentralized trust setting.
    A successful process of secure aggregation relies on a transparent consensus among parties, in which each party can present its `decision' or `opinion' on a participation matrix that reflects its role in each secure aggregation round and its expectation of the proportion of benign parties in the FL training.
    \textit{DeTrust-FL} utilizes decentralized functional encryption to perform the fundamental secure computing task. 
    
    \item \textit{DeTrust-FL} addresses the lack of transparency that enables isolation attacks during secure aggregation by assuring that parties' model updates are included in the aggregated model in a private and secure manner. \textit{DeTrust-FL} also mitigates privacy leaks caused by recently demonstrated disaggregation attacks. 
    To this end, we design a distributed consensus trust setting, which at the same time can address potential attacks where a curious aggregator colluding with a small number of  parties. 
    
    \item 
    We describe an implementation of \textit{DeTrust-FL} and demonstrate its applicability to train neural networks. According to our experimental evaluation, conducted in a cloud-based distributed setup,  \textit{DeTrust-FL} provides the state-of-the-art model accuracy, communication and computation efficiency, while protecting the privacy of model updates from each party.
\end{itemize}
To the best of our knowledge, this is the first PPFL solution that gives each party fine-grained control over how their model updates are included and whether or not their participation matrix leaks private information, hence increasing the party's trust on collaborating via PPFL framework.

\section{Motivation and Preliminaries}
\label{sec:motivation-preliminaries}
The goal of SMC schemes is to protect the privacy of the inputs. In the FL setting, this means ensuring that nothing is revealed by the aggregation process other than the aggregated model.
Attacks that compromise the privacy of existing SMC schemes stem from the ability of the aggregator to  \textit{i)} analyze the log of aggregated results or \textit{ii)} manipulate the data that is fed to the SMC procedure.

A curious aggregator may launch a \textit{disaggregation attack} by analyzing the outputs of the SMC over the training rounds. A matrix containing these outputs across training rounds and the knowledge of what parties participated in each round is used to build a system of equations that can be used to solve for the outputs of one of the parties. Note that this curious aggregator does not deviate from the training process and simply analyzes the matrix.
To prevent this attack, it is necessary to ensure that the resulting matrix does not enable the construction of equations that can be solved to isolate a single input. 
Existing techniques \cite{lam2021gradient,so2021securing} only work for very large federations and are not appropriate for consortia with a small number of participants.

\textit{Isolation attacks} consist of 
Isolation attacks consist of selectively inputting model updates to the SMC process in such a way that certain model updates can be isolated. Isolation attacks violate privacy and may lead to biased models.
Techniques to prevent this attack have been proposed in \cite{xu2019hybridalpha} where an \textit{inference prevention module} is introduced to prevent potential malicious behaviors during the secure aggregation procedure; however, this module is deployed by a fully trusted authority which is difficult to embody in real-life scenarios. 

Finally, \textit{replay attacks} occur when a malicious aggregator decides to send \textit{old} model updates to the SMC process to force the construction of a vulnerable matrix. Both isolation and replay attacks are active attacks and are very difficult to detect. A potential solution is using specialize trusted hardware that is typically expensive and unavailable to federations.
The goal of this paper is to enable SMC while preventing disaggregation, isolation and replay attacks without requiring specialize hardware or a \textit{third party authority} (TPA).

\vspace{2pt}
\noindent\textbf{Preliminaries of Decentralized Functional Encryption:}
Functional encryption (FE) is a family of cryptosystems that allows the computation of functions over encrypted data and as a result the final function value is obtained in plaintext~\cite{boneh2011functional}. FE schemes have shown promising efficiency improvements to address the secure multi-party aggregation challenge compared to  existing homomorphic encryption-based approaches \cite{xu2019hybridalpha}. 
In \cite{xu2019hybridalpha}, FL systems made use of a functional encryption scheme \cite{abdalla2018multi} that enabled the computation of inner product using a TPA.
Unlike the functional encryption scheme used in \cite{xu2019hybridalpha}, this paper uses the recently proposed decentralized multi-client functional encryption (DMCFE) schemes \cite{chotard2018decentralized, abdalla2019decentralizing, chotard2020dynamic}.

DMCFE introduces two very interesting features: \textit{i)} each party has a partial key and it does not require a TPA to generate functional keys for all the federation, and \textit{ii)} it ensures that a particular functional key can only be used to compute a particular ciphertext by cryptographically marking the ciphertext with a \textit{label}.
Let $\mathcal{F}=\{\mathcal{F}_{\rho}\}_{\rho}$ be a family (indexed by $\rho$) of sets $\mathcal{F}_{\rho}$ of functions $f:\mathcal{X}_{\rho,1} \times \cdots \times \mathcal{X}_{\rho,n_{\rho}} \to \mathcal{Y}_{\rho}$ and $\mathcal{L}=\{0,1\}^{*} \cup \{\perp\}$ be a set of labels. 
A DMCFE scheme for the function family $\mathcal{F}$ and the label set $\mathcal{L}$ is a tuple of six algorithms $\mathcal{E}_{\mathtt{DMCFE}}=(\mathtt{Setup}, \mathtt{KeyGen}, \mathtt{KeyDerShare}, \mathtt{KeyDerComb}, \mathtt{Enc}, \mathtt{Dec})$:
\begin{itemize}
    \item[-] $\mathtt{Setup}(1^{\lambda}, 1^{n})$ : 
    Takes as input a security parameter $\lambda$ and the number of parties $n$, and generates public parameters $\mathtt{pp}$.
    \item[-] $\mathtt{KeyGen}(\mathtt{pp})$: 
    Takes as input the public parameters $\mathtt{pp}$ and outputs $n$ secret keys $\{\mathtt{sk}_{i}\}_{i\in[n]}$.
    \item[-] $\mathtt{KeyDerShare}(\mathtt{pp}, \mathtt{sk}_{i}, f)$: 
    Takes as input the public parameters $\mathtt{pp}$, a secret key $\mathtt{sk}_{i}$ and a function $f\in\mathcal{F}_{\rho}$, and outputs a partial functional decryption key $\mathtt{dk}_{i}$.
    \item[-] $\mathtt{KeyDerComb}(\mathtt{pp}, \{\mathtt{dk}_{i}\}_{i\in[n]})$:
    Takes as input the public parameters $\mathtt{pp}$, $n$ partial functional decryption keys $\{\mathtt{dk}_{i}\}$ and outputs the functional decryption key $\mathtt{dk}_{f}$.
    \item[-] $\mathtt{Enc}(\mathtt{pp}, \mathtt{sk}_{i}, x_i, l)$: 
    Takes as input the public parameters $\mathtt{pp}$, a secret key $\mathtt{sk}_{i}$, a message $x_i\in\mathcal{X}_{\rho,i}$ to encrypt, a label $l\in\mathcal{L}$, and outputs ciphertext $\mathtt{ct}_{i,l}$. 
    \item[-] $\mathtt{Dec}(\mathtt{pp}, \mathtt{dk}_{f}, \{\mathtt{ct}_{i,l}\}_{i\in[n]})$: 
    Takes as input the public parameters $\mathtt{pp}$, a functional key $\mathtt{dk}_{f}$ and $n$ ciphertexts under the same label $l$ and outputs a value $y\in\mathcal{Y}_{\rho}$.
\end{itemize}

\section{The Proposed \textit{DeTrust-FL} Framework}
\label{sec:ht}


\textit{Threat Model and Assumptions}.
The objective of \textit{DeTrust} is to ensure that the FL training process can provide privacy of the \textit{input}, which guarantees the privacy of each party's local model, and the \textit{output}, which ensures the privacy of the aggregated global model.
The input privacy is achieved via secure aggregation, which allows the aggregator to learn only the aggregated model, and the output privacy is achieved through differential privacy.
We assume that a malicious aggregator may attempt to infer private information of a target party during or after the training process. 
We focus our attention on disaggregation, isolation, and replay attacks described in Sec.~\ref{sec:motivation-preliminaries}, and refer to the aggregator as \textit{semi-trusted}.
Moreover, we assume that the aggregator may collude with \textit{a subset of parties} of bounded size
to acquire private information of the other parties. 
We assume that parties are honest-but-curious, and do not consider attacks where malicious parties manipulate the model updates to poison the ML model or reduce its accuracy or to create a denial of service attack.


\begin{figure}
    \centering
    \includegraphics[width=\textwidth]{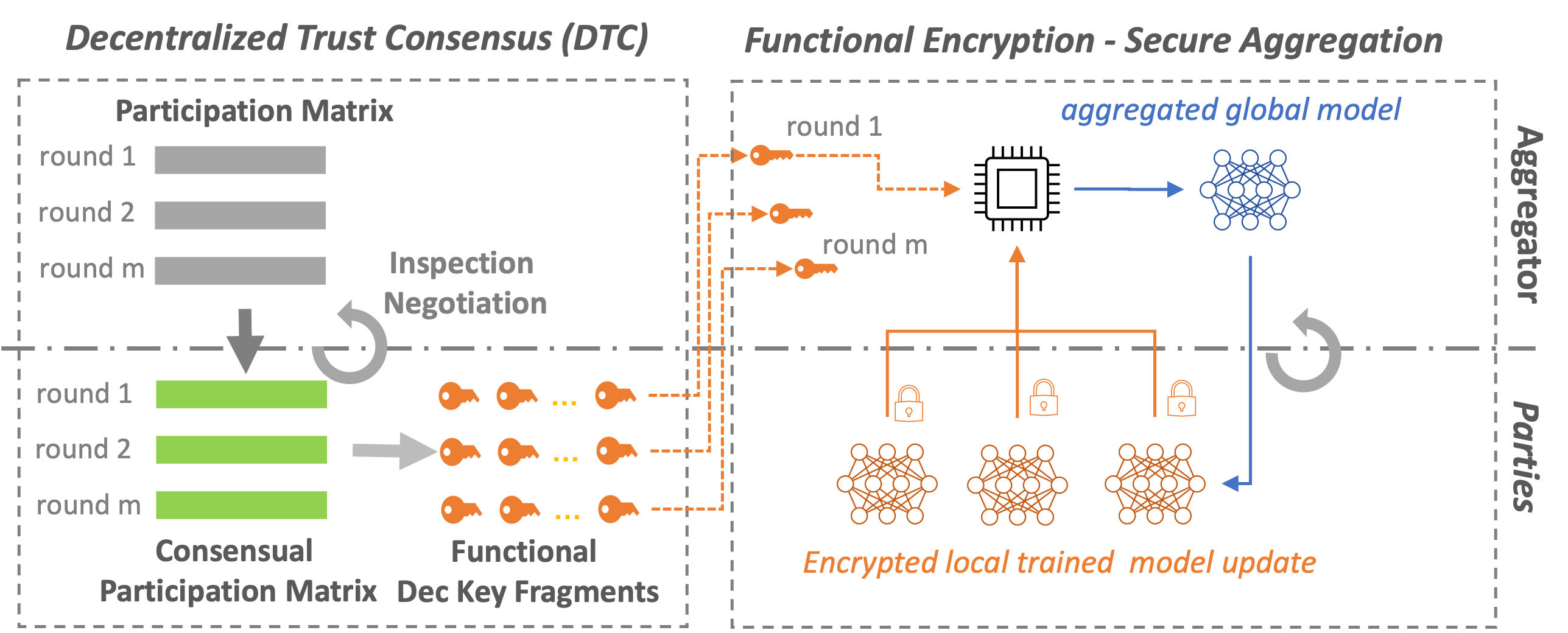}
    \caption{Overview of \textit{DeTrust-FL} framework.}
    \label{fig:overview}
\end{figure}

\textit{How does DeTrust-FL work?}
Figure \ref{fig:overview} depicts the overview of \textit{DeTrust-FL}.
Similar to \cite{liu2019secure, zhang2020batchcrypt,xu2019hybridalpha}, \textit{DeTrust-FL} adopts advanced cryptography-based secure aggregation to achieve privacy-preserving FL coupled with a consensus mechanism
that provides transparency to the secure aggregation process.

Before training starts, \textit{DeTrust-FL} performs a decentralized consensus through its decentralized trust consensus (DTC) module.
The DTC module enables each party to inspect and negotiate the \textit{participation matrix} that
dictates what parties model updates will be input into the SMC process and with what weights during each round of training. Initially, the aggregator proposes a participation matrix to determine which parties will be selected for each training round. Parties engage in a protocol that enables them to agree on a matrix that does not allow for inference.
At the end of DTC process, each party agrees on an updated participation matrix and uses it to generate the functional decryption key fragments, which represent its trust on the FL training process.
Training can only proceed after the successful completion of the DTC.

At each training round, each party trains a local model, encrypts the model update using the DMCFE cryptosystem \cite{abdalla2019decentralizing}, and sends the encrypted model updates to the aggregator.
As part of the encryption, each party \textit{cryptographically labels} its model update with the current training round.
The aggregator then computes the new \textit{global model} by aggregating all encrypted inputs. 
The aggregator can execute secure aggregation by decrypting the set of encrypted model updates if and only if it receives all functional key fragments from parties for each training round. 
In addition, for the decryption to be successful, the aggregator can use the cryptographic key only for the current round, as the encrypted model updates are labeled with that round.
Once the aggregator performs the decryption, it returns the new global model to each party.


Importantly, while our \textit{DeTrust-FL} solution utilizes the same functional encryption family as previous PPFL approach \cite{xu2019hybridalpha}, it does not introduce any additional entities (e.g., the online trusted third-party authority suggested in \cite{xu2019hybridalpha}).
Thus, \textit{DeTrust-FL} is architecturally equivalent to the standard FL paradigm \cite{konevcny2016federated}.
Additionally, \textit{DeTrust-FL} supports both \textit{average} and \textit{weighted} fusion methods directly, whereas the majority of secure aggregation approaches support just the former. This is discussed in Section~\ref{sec:comparison-with-existing}.
In the following, we present the notation we use and more details of the private training process.




\subsection{Notation}
\label{sec:notation}
We use the following notation throughout the rest of the paper.
Let $\mathcal{A}$ be the aggregator and $\mathcal{S}_{\mathcal{P}}$ be the set of $n$ parties that participate in $m$ FL training rounds. Let $\mathcal{D}_j$ denote the dataset of party $\mathcal{P}_j\in\mathcal{S}_{\mathcal{P}}$.
Let $\mathcal{M}_{G}^{(i)}$ and $\mathcal{M}_{\mathcal{P}_{j}}^{(i)}$ denote the global model and party $\mathcal{P}_j$'s local model, respectively, in FL round $i$. 
We use $\pmb{V} \in \mathbb{R}^{m\times n}$ to denote the \textit{participation matrix}, where each row $\pmb{V}_{i}=(v_{i,\mathcal{P}_{0}}, ..., v_{i, \mathcal{P}_{n}})_{1\le i\le m}$ represents the fusion vector for $i$-th training round, defining 
the weights assigned to parties' local models for aggregation. That is, $\mathcal{M}^{(i)}_{G}=\sum_{j}v_{i, \mathcal{P}_{j}}\mathcal{M}^{(i)}_{\mathcal{P}_{j}}$.
The column $\pmb{V}_{,\mathcal{P}_{j}}$ represents the fusion weights of $\mathcal{P}_{j}$ for all $m$ training rounds.
Note that the entry $V_{i,j}=0$ means that during round $i$ the local model of party $j$ will not be added to the model.
Let $DP_{SMC}(\epsilon, \mathcal{M}^{(i)}_{\mathcal{P}_{i}})$ denote 
the DP mechanism proposed in~\cite{truex2019hybrid}, which reduces the amount of injected noise  by leveraging secure aggregation.
We use $t_{g}$ 
to denote the minimum number of parties that need to be aggregated in a round.

\subsection{Training Process in \textit{DeTrust-FL}}
\label{sec:ht:framework}


\textit{DeTrust-FL} operation relies in two primary steps: \textit{reaching a trust consensus} on a participation matrix that dictates who will participate during a particular round of training and what weights will be used to aggregate participant's model updates, and then, \textit{performing privacy-preserving training} among parties in a decentralized setting. 

\noindent\textbf{Decentralized Trust Consensus (DTC)}.
The objective of DTC is to establish an agreed-upon participation matrix $\pmb{V}$ for $m$ rounds of FL training among $n$ parties (see Sec.~\ref{sec:notation} for the definition).
The \textit{DTC} process begins with the aggregator proposing an initial participation matrix $\pmb{V}$. 
An honest aggregator may propose a matrix that meets a batch partitioning (BP) constraint to prevent inference \cite{so2021securing} and the global threshold requirements. 
A malicious aggregator may provide other input; however, the following steps would prevent the attack. 

Each party investigates the proposed $\pmb{V}$ as follows:
(i) does $\pmb{V}$ violate the BP criterion?
(ii) does $\pmb{V}_{,\mathcal{P}_{j}}$ accurately reflect fusion weight of party $\mathcal{P}_j$ for each round of FL training and (ii) does $\pmb{V}$ ensure that a minimum number of parties are included into secure aggregation for all rounds?
If any of these inspections fail, party $\mathcal{P}_j$ can propose a new participation vector $\pmb{V}_{,\mathcal{P}_{j}}$.
Then, $\mathcal{A}$ compiles all received assessments and conducts another round of party inspections until all the parties agree on the updated $\pmb{V}$.
Finally, each party generates functional decryption key fragments $\mathtt{dk}^{\pmb{V}} \in \mathbb{Z}^{m\times 1}$ based on the previously negotiated $\pmb{V}$ using parties' secret key.
We will formalize these procedures in Section~\ref{sec:ht:dtc}.

\noindent\textbf{Privacy-Preserving Training}.
For each training round, the aggregator queries parties with the current global model.
Each party verifies it is participating in the training process during the current training round, which we refer to as being \textit{enrolled} in a round of training, according to the previously agreed participation matrix.
The enrolled party first trains a local model update using the current global model and its local dataset.
In DeTrust-FL, we use the same hybrid method adding differential privacy noise as proposed in \cite{truex2019hybrid} to ensure both input and output privacy.
Following that, the aggregator waits and collects a sufficient number of protected models in accordance with the agreed participation matrix.
Then, using the collection of key fragments corresponding to current round's fusion weight vector $\pmb{V}_{i}$, the aggregator recovers the functional decryption key $\mathtt{dk}^{\pmb{V}_{i}}$ for the round.
The recovered $\mathtt{dk}^{\pmb{V}_{i}}$ is also bound to a specific round index and cannot be used in any other training round, as the DMCFE cryptosystem \cite{chotard2018decentralized, abdalla2019decentralizing} guarantees.
Finally, the aggregator obtains the new global model by decrypting the collection of encrypted model updates without learning anything from each of the encrypted model updates.
Notice that $\mathtt{dk}^{\pmb{V}_{i}}$ is generated collaboratively by the parties based on the trusted participation matrix that is negotiated through the DTC procedure.

\begin{algorithm}[t]
    \SetAlgoLined
    \caption{Privacy-Preserving \textit{DeTrust-FL} Training}
    \label{alg:framework}
    
    \SetKwInput{KwInput}{Input}                
    \SetKwInput{KwOutput}{Output}              
    \SetKwProg{Fn}{function}{}{}
    \SetKwRepeat{Do}{do}{while}%
    
    \small
    \KwInput{Party set $\mathcal{S}_{\mathcal{P}}$, each party $P_j$ has its own dataset $\mathcal{D}_{{j}}$.}
    \KwOutput{Final trained global model $\mathcal{M}_{G}^{(m)}$.}
    
    Each entity initializes cryptographic keys\;
    Perform \textit{Decentralized-Trust-Consensus} to build $\pmb{V}$ and $\mathtt{dk}^{\pmb{V}}$\;
    Aggregator $\mathcal{A}$ initializes model $\mathcal{M}_{G}^{(0)}$\;
    \ForEach{round ${i} \in \{1, ..., m\}$ }{
        \ForEach{$\mathcal{P}_{j} \in \mathcal{S}_{\mathcal{P}}$}{
            $\mathcal{A}$ queries $\mathtt{msg}_{\mathtt{req}}=(\mathcal{M}^{(i-1)}_{G}$, hyperparameters)\;
            \If{$\mathcal{P}_{j}$ is enrolled in $\pmb{V}_{i}$}{
                $\mathcal{M}^{(i)}_{\mathcal{P}_{j}} \gets$ 
                $train(\mathcal{D}_{{j}}, \mathcal{M}^{(i-1)}_{G})$\;
                $[\mathcal{M}^{(i)}_{\mathcal{P}_{j}}] \gets \mathcal{E}_{\mathtt{DMCFE}}.\texttt{Enc}_{\texttt{sk}_{\mathcal{P}_{j}}}(\texttt{DP}_{SMC}(\epsilon, \mathcal{M}^{(i)}_{\mathcal{P}_{j}}))$\;
                $\mathcal{P}_{j}$ replies $[\mathcal{M}^{(i)}_{\mathcal{P}_{j}}]$ to $\mathcal{A}$\;
            }
        }
        $\mathcal{S}_{\mathtt{msg}_{\mathtt{resp}}} \gets \mathcal{A}$ waits and collects responses\; 
        \If{response status $[\mathcal{M}^{(i)}_{\mathcal{P}_{j}}] \in \mathcal{S}_{\mathtt{msg}_{\mathtt{resp}}}$ satisfies $\pmb{V}_{i}$}{
            $\mathcal{A}$ recovers $\texttt{dk}_{i}$ from fragments $\mathtt{dk}^{\pmb{V}}_{i}=\{\mathtt{dk}_{i, \mathcal{P}_{j}}\}$\;
            $\mathcal{M}^{'}_{G} \gets \mathcal{E}_{\mathtt{DMCFE}}.\texttt{Dec}_{\texttt{dk}_{i}}(\{[\mathcal{M}^{(i)}_{\mathcal{P}_{i}}]\})$\;
            $\mathcal{A}$ updates $\mathcal{M}^{(i)}_{G} \gets \mathcal{M}^{'}_{G}$\;
        }
    }
    \Return final model $\mathcal{M}_{G}^{(m)}$\;
\end{algorithm}

\noindent\textbf{Algorithm Overview}.
Algorithm~\ref{alg:framework} gives an overview of how \textit{DeTrust-FL} works.
Each participant creates its cryptographic keys (Line 1).
To initiate the learning process, $\mathcal{A}$ executes the \textit{Decentralized-Trust-Consensus} function with all parties to agree on a participation matrix $\pmb{V}$ and corresponding functional decryption key fragments $\mathtt{dk}^{\pmb{V}}$ (Line 2).
We present the decentralized trust consensus in detail in Section~\ref{sec:ht:dtc}.
Then, $\mathcal{A}$ initiates the training process in the same manner as standard FL does by sending a query message to all parties (Lines 4-6).
Each enrolled party trains its local model and adds differential private noise to its model (Lines 7-10).
When all responses are received, $\mathcal{A}$ needs to check the response status $\mathcal{S}_{\mathtt{msg}_{\mathtt{resp}}}$ to ensure the required relies are satisfied according to the agreed-upon participation of current training round $\pmb{V}_{i}$ (Line 12). 
If the quorum verification passes, $\mathcal{A}$ recovers current round's decryption key $\mathtt{dk}^{\pmb{V}_{i}}$ and performs the decryption among the encrypted model collection to obtain the new global model (Lines 13-15).
We elaborate on the hybrid methodology used in privacy-preserving training in Section~\ref{sec:comparison-with-existing}.

\subsection{Decentralized Trust Consensus Process}
\label{sec:ht:dtc}
 

As we have illustrated, $\pmb{V}_{i}$ defines the assigned fusion weight used for each party's local model at $i$-th round and functional decryption key $\mathtt{dk}^{\pmb{V}}_{i}$ is collaboratively generated based on the provided $\pmb{V}_{i}$. 
The aggregator can decrypt collected encrypted local models to obtain the aggregated global model if and only if (i) the aggregator proposed a ``proper'' $\pmb{V}$ that is unanimously agreed among parties, and (ii) those parties collaboratively generate  $\mathtt{dk}^{\pmb{V}}$ key.
A mutually agreed-upon $\pmb{V}$ can represent the parties' unanimity of trust in the secure aggregation process, as well as in the privacy-preserving FL training.
Thus, achieving such an agreement is the major objective of decentralized trust consensus setting, as various parties may have varying degrees of trust in a FL collaboration.

\begin{algorithm}[h]
    \SetAlgoLined
    \caption{Decentralized Trust Consensus}
    \label{alg:dtc}
    
    \SetKwInput{KwInput}{Input}                
    \SetKwInput{KwOutput}{Output}              
    \SetKwProg{Fn}{function}{}{}
    \SetKwRepeat{Do}{do}{while}%
    
    \small
    \KwInput{$n$:= num. parties and $m$:= num. training rounds.}
    \KwOutput{Agreed-upon $\pmb{V}$ and decryption key fragments $\texttt{dk}^{\pmb{V}}$.}
    
    $\mathcal{P}_{j}\in\mathcal{S}_{\mathcal{P}}$ initializes  and sends local trust threshold $t_{\mathcal{P}_{j}}$\;
    $\mathcal{A}$ collects and sets up the agreed-upon trust threshold $t_{g}=\mathtt{max}(\{t_{\mathcal{P}_{j}}\}_{\mathcal{P}_{j}\in\mathcal{S}_{\mathcal{P}}})$\;
    $\mathcal{A}$ initializes $\pmb{V'}^{m\times n}$ s.t. $\vdash\texttt{BP}(\pmb{V'}) \wedge \forall i, \|\pmb{V'}_{i}\|_{\ne 0} \ge t_g$\;
    \Fn{Decentralized-Trust-Consensus($t_{g}, m, \mathcal{S}_{\mathcal{P}}$)}{
        $\mathcal{A}$ initializes empty $\texttt{dk}^{\pmb{V}} \in \pmb{0}^{m\times n}$ and threshold set $\mathcal{S}_{t}$\; 
        \Do{all entities do not agree on participation matrix $\pmb{V}$}{
            $\mathcal{A}$ queries all parties with $\pmb{V'}$\;
            \ForEach{$\mathcal{P}_{j} \in \mathcal{S}_{\mathcal{P}}$}{
                $\mathtt{vf}_{\mathcal{P}_{j}} \gets$ \textit{Party-Inspect-Negotiate}($\pmb{V'}_{BP}, t_g$)\;
                \If{$\mathtt{vf}_{\mathcal{P}_{j}}$ $=\pmb{V}^{\textit{suggest}}_{, \mathcal{P}_{j}}$ is suggested}{
                    \lIf{$\mathcal{A}$ accepts $\pmb{V}^{\textit{suggest}}_{, \mathcal{P}_{j}}$}{
                        $\pmb{V'}_{,j} \gets \pmb{V}^{\textit{suggest}}_{, \mathcal{P}_{j}}$}
                    \lElse{$\pmb{V'}_{,j} \gets \pmb{0}^{m\times1}$}
                }
            }
            $\mathcal{A}$ updates $\pmb{V'}$ s.t. $\vdash\texttt{BP}(\pmb{V'}) \wedge \forall i, \|\pmb{V'}_{i}\|_{\ne 0} \ge t_{g}$\;
        }
        \ForEach{$\mathcal{P}_{j} \in \mathcal{S}_{\mathcal{P}}$}{
            \ForEach{row $\pmb{V}_i \in \pmb{V}$}{
                $\texttt{dk}_{i, \mathcal{P}_{j}} \gets \mathcal{E}_{\mathtt{DMCFE}}.\texttt{KeyDerShare}_{\texttt{sk}_{\mathcal{P}_{j}}}(\pmb{V}_{i})$\;
            }
            $\mathcal{P}_{j}$ replies to $\mathcal{A}$ with $\{\texttt{dk}_{i, \mathcal{P}_{j}}\}_{i\in\{1...m\}}$\;
            $\mathcal{A}$ fills column $\texttt{dk}^{\pmb{V}}_{, j} \gets \{\texttt{dk}_{i, \mathcal{P}_{j}}\}_{i\in\{1...m\}}$\;
        }
        \Return $\pmb{V}$ and $\texttt{dk}^{\pmb{V}}$\;
    }
    
    \Fn{Party-Inspect-Negotiate( $\pmb{V}^{'}, t_g$)}{
        $\mathcal{P}_{j}$ initializes $\mathtt{vf}_{\mathcal{P}_{i}} = \{\}$\;
        \lIf{$t_g < t_{\mathcal{P}_{j}}$}{\Return \texttt{refuse}}
        \lIf{\textbf{not} $\vdash\texttt{BP}(\pmb{V'})$}{\Return \texttt{violate-BP}}
        \ForEach{row $i \in \pmb{V}^{'}$}{
            \If{$\|\pmb{V}_{i}\|_{\ne 0}  \ge t_{g}$}{
                \lIf{$\pmb{V}_{i,\mathcal{P}_{j}}$ is expected }{$\mathtt{vf}_{\mathcal{P}_{j}} \gets$ \texttt{accept}}
                \lElse{$\mathtt{vf}_{\mathcal{P}_{j}} \gets \pmb{V}^{\textit{suggest}}_{i,\mathcal{P}_{j}}$}
            }
            \lElse{$\mathtt{vf}_{\mathcal{P}_{j}} \gets$ \texttt{refuse}}   
        }
        \Return $\mathtt{vf}_{\mathcal{P}_{j}}$
    }
\end{algorithm}

\noindent\textbf{Detailed Procedure}.
Algorithm~\ref{alg:dtc} demonstrates how parties can agree on $\pmb{V}$ and then collaboratively generate $\mathtt{dk}^{\pmb{V}}$ key fragments for secure aggregation.
Prior to the DTC process, \textit{DeTrust-FL} ensures that each party is initialized with a threshold $t_{\mathcal{P}_{j}}$ 
that defines the individual trust threshold of the benign party proportion among $n$ parties.
Next, $\mathcal{A}$ collects parties' trust threshold and sets up the global trust threshold as $t_{g}=\mathtt{max}(\{t_{\mathcal{P}_{j}}\}_{\mathcal{P}_{j}\in\mathcal{S}_{\mathcal{P}}})$ (Lines 1-2).
The DTC process begins with $\mathcal{A}$ providing an initial $\pmb{V'}^{m\times n} \in \mathbb{R}$ that complies with the Batch Partitioning (BP) criteria \cite{so2021securing} and $\forall i, \|\pmb{V'}_{i}\|_{\ne 0} \ge t_g$ (Line 3).
The BP verification ensures that the participation of the parties in a particular training round do not enable disaggregation attacks.
Next, $\mathcal{A}$ repeatedly queries all parties to obtain inputs on $\pmb{V'}$, updates expected threshold $t_{\max}$, re-proposes $\pmb{V'}$, until all parties reach to an agreed $\pmb{V}$ (Lines 6-14).
To effectively re-propose a participation matrix $\pmb{V}$, $\mathcal{A}$ acquires suggested fusion weight $\pmb{V}^{\textit{suggest}}_{, \mathcal{P}_{j}}$ from party $\mathcal{P}_{j}$ to form new $\pmb{V}$ and ensure that:
\begin{equation*}
    \left\{
    \begin{array}{ll}
        \text{comply with BP criteria, i.e., } \vdash\texttt{BP}(\pmb{V})\\
        \forall i \in \{1,...,m\}, \|\pmb{V}_{i}\|_{\ne 0} \ge t_{g} \;i.e.\; t_{g} = \max(\{t_{\mathcal{P}_{j}}\})\\
        \forall j \in \{1,...,n\},  \pmb{V}_{,\mathcal{P}_{j}} \text{ is accepted by } \mathcal{P}_{j}\\
    \end{array}
    \right.
\end{equation*}

When a party receives a participation matrix query message, each party first verifies the global trust threshold and the BP criteria compliance (Lines 23-24).
For each row of compliance $\pmb{V}$, $\mathcal{P}_{j}$ inspects the enrollment status, trust threshold, and expected fusion weight that are associated to itself.
If all inspections are successful, the party can respond with an accept signal; otherwise, it may respond with corresponding feedback (Lines 26-30).
Finally, all parties agree on a participation matrix $\pmb{V}$ and generate functional decryption key fragments $\{\texttt{dk}_{i, \mathcal{P}_{j}}\}_{i\in\{1...m\}}$ corresponding to that matrix using each party's private key (Lines 15-20).

\subsection{Contrasting \textit{DeTrust} with Existing Techniques}
\label{sec:comparison-with-existing}

\noindent\textbf{Decentralizing Trust}.
\textit{DeTrust-FL} relies on a novel FE cryptosystem that does not rely on a fully trusted entity
overcoming some of the adoption problems of previously proposed FE-based systems including \cite{xu2019hybridalpha},
where $\texttt{dk}$ key is generated by a third-party entity, and the inference prevention module is also controlled by it.
As a result, one limitation of \cite{xu2019hybridalpha} is that all parties needs to fully trust the third-party entity. 
\textit{DeTrust-FL} completely removes this limitation. In \textit{DeTrust-FL}, $\texttt{dk}$ key is collaboratively generated by parties.
Each party has the ability to verify the secure aggregation process in a fine-grained manner by controlling the partial functional decryption key generation.
Without enough $\texttt{dk}$ key fragments, the aggregator cannot decrypt to acquire the aggregated global model.
Only if the totality of the parties reach consensus on the fusion weights, the aggregator can perform the secure aggregation process.
In such a design, \textit{DeTrust-FL} decentralizes the trust reliance on a centralized crypto dealer.

\noindent\textbf{Transparency over the Secure Aggregation Process}.
Secure aggregation techniques suffer from lack of transparency with respect to the inputs an aggregator
feeds to the SMC procedure.
In \textit{DeTrust-FL}, we allow $\mathcal{P}_{j}$ to 
(i) define their personalized local trust threshold $t_{\mathcal{P}_{j}}$ for $\pmb{V}$, 
and (ii) verify the number of parties participating in a particular round of training and
the individual weights used to aggregate them $\pmb{V}_{,\mathcal{P}_{j}}$ in $\pmb{V}$.
The former setting represents the party's confidence in enrolled parties for training process, whereas the local trust threshold $t_{\mathcal{P}_{j}}$ indicates the party's expectation on minimum percentage of enrolled benign parties for one training round.
For the enrolled training rounds it allows the aggregator and the party to negotiate on the excepted fusion weights used in the FL training.
Because the matrix is inextricably linked to the functional decryption key fragments, it dictates the results of the secure aggregation.
This design removes the assumption of a trusted and honest aggregator.



\noindent\textbf{Hybrid Methodology Compatibility}.
The secure aggregation of \textit{DeTrust-FL} is achieved using DMCFE technique, in which each party encrypts its local model update using a secret key $\mathtt{sk}_{\mathcal{P}_{j}}$ and the aggregator is issued a functional decryption key $\mathtt{dk}_{i}$ to decrypt the set of encrypted local models in order to obtain the aggregated model directly without learning the protected local models at $i$-th training round, as illustrated in Line 10 and Line 15 of Algorithm~\ref{alg:framework}.
After obtaining the aggregated global model, the aggregation can conduct another round of FL training.
The main distinction between our secure aggregation solution and other FE-based secure aggregation solutions \cite{xu2019hybridalpha} is that in \textit{DeTrust-FL}, $\mathtt{dk}_{i}$ is recovered from a set of key fragments generated by parties in accordance with the agreed-upon participation matrix, rather than being issued by a centralized trusted authority \cite{xu2019hybridalpha}.
Additionally, our secure aggregation can be used with the hybrid methodology described in \cite{truex2019hybrid}, in which we first inject a certain amount of differential privacy noise into the locally trained model and then encrypt the noise-injected model achieving higher model performance.
As a result, \textit{DeTrust-FL} retains the benefit of noise reduction for the global model through secure aggregation approaches.

\noindent\textbf{Support for Fusion Algorithms}.
\textit{DeTrust-FL} supports two commonly used algorithms to aggregate model updates:
the \textit{average fusion method} that averages aggregated local models and the \textit{weighted fusion method}  that fuses local models from each party with assigned party weight, e.g., FedAvg. 
Note that the latter is not supported in emerging additive HE-based (e.g., Paillier) secure aggregation solutions \cite{liu2019secure, zhang2020batchcrypt}, threshold Pailler \cite{truex2019hybrid} based secure aggregation, and pairwise masking based protocol \cite{bonawitz2017practical}.
As explained above, $\pmb{V}_{i}$ specifies the fusion weight assigned to each party's local model in the aggregated global model at the $i$-th round.
In the average fusion method, $\pmb{V}_{i}$ is set as an all-$\frac{1}{\|\pmb{V}_{i}\|_{\ne 0}}$ vector in case that all parties are enrolled in the FL training.
\textit{DeTrust-FL} also supports FedAvg by ensuring that the weights in matrix $V$ reflect the number of samples in each party.

\section{Security and Privacy Analysis}
\label{sec:tpsa}

\subsection{Security of Cryptographic Infrastructure}
As DMCFE is the underlying cryptographic infrastructure of our secure aggregation approach, the security of DMCFE is critical in the \textit{DeTrust-FL} framework.
Our adoption does not change any algorithms of the DMCFE scheme.
According to formal security proof presented in \cite{abdalla2019decentralizing}, the encrypted model in \textit{DeTrust-FL} has ciphertext indistinguishability and is secure against adaptive corruptions under the classical DDH assumption.
To minimize redundancy, we will not discuss the correctness and security proofs to DMCFE here, and readers can refer to \cite{abdalla2019decentralizing} for more details.

\subsection{Privacy Guarantee}
In this section, we give an overview of the privacy guarantees for \textit{DeTrust-FL}.

\noindent\textbf{Privacy of the Aggregated Global Model}.
Similarly to the comparable privacy-preserving federated learning proposals \cite{xu2019hybridalpha, truex2019hybrid}, \textit{DeTrust-FL} uses a differential privacy mechanism that takes advantage of the secure aggregation procedure to protect the privacy of the output, i.e., the aggregated global model (in each round).
Following the theoretical analysis reported in \cite{truex2019hybrid}, 
the output model of \textit{DeTrust-FL} also achieves the same differential privacy guarantee.

\noindent\textbf{Privacy of Each Party's Local Model}.
In the threat model, we consider a \textit{semi-trusted aggregator} $\mathcal{A}$ and a \textit{limited number of colluding parties}, with disaggregation, isolation, and replay attacks described in Sec.~\ref{sec:motivation-preliminaries}.

\noindent\textit{Inference Attack I (Isolation attack without collusion):} $\mathcal{A}$ can launch an inference attack targeting a specific party by manipulating the participation matrix for target $i$-th training round at target party $\mathcal{P}_{j}$.
In the case of average fusion method, instead of setting up an all-one vector 
$\pmb{V}_{i}$, $\mathcal{A}$ may set $\pmb{V}_{i}$ as $(0,.., v_{i,\mathcal{P}_{j}},...,0), v_{i,\mathcal{P}_{j}}=1$. 
Thus, $\sum_{i} \pmb{V}_{i}\mathcal{M}_{i}=\mathcal{M}_{\mathcal{P}_{j}}$.
As a consequence, $\mathcal{A}$ can infer the input of $\mathcal{P}_{j}$.
Our  proposed \textit{inspection} module filters malicious $\pmb{V}_{i}$ by counting non-zero elements as demonstrated in Section~\ref{sec:ht:dtc}, which prevents isolation attacks. 

\noindent\textit{Inference Attack I (Isolation Attack with Colluding Parties):} To highlight this attack, consider $n=7$ parties. Now, $\mathcal{A}$ may propose a malicious $\pmb{V}_{i}=$ (1,1,1,0,0,0,0) as one row of participation matrix to acquire the functional decryption key fragments. 
If $\mathcal{A}$ can collude with two parties, then it can infer a benign party's input. 
Our  proposed \textit{inspection} module ensures that in each round at least $t_g$ parties' models are included in the secure aggregation. Therefore, as long as the the aggregator colludes with fewer than $t_g-2$ parties, it cannot infer the local model of any individual party. 

\noindent\textit{Inference Attack III (Disaggregation Attack):} $\mathcal{A}$ can launch the \textit{disaggregation attack} over multiple rounds of FL training by exploiting the side channel information (i.e., participation matrix), as demonstrated in \cite{so2021securing, lam2021gradient}. 
As illustrated in Section~\ref{sec:ht:dtc}, the \textit{DeTrust-FL} design allows each party to examine the participation matrix to determine whether it complies with the Batch Partitioning (BP) criteria, a defense approach demonstrated in \cite{so2021securing}. To prevent such a multi-round disaggregation attack, \textit{DeTrust-FL} incorporates this BP criterion into our secure aggregation in a decentralized setting.

\noindent\textit{Inference Attack IV (Replay Attack):} To highlight this attack, suppose $n=6$ and let $\pmb{V}_{i_1} = (1, 1, 1, 1, 0, 0)$ and $\pmb{V}_{i_2} = (1, 1, 1, 1, 0, 0)$ for some $i_2 > i_1$. In other words, in rounds $i_1$ and $i_2$, model updates from parties 1 through 4 will be averaged together. Now, the aggregator can target a specific party, say $\mathcal{P}_1$, by launching a replay attack in round $i_2$ by using ciphertexts from round $i_1$ for parties $\mathcal{P}_2$, $\mathcal{P}_3$, and $\mathcal{P}_4$. Note that in round $i_1$ secure aggregation yields $\sum_{j=1}^{4}\mathcal{M}_{\mathcal{P}_j}^{(i_1)}$. Now, due to the replay attack in round $i_2$, secure aggregation will yield $\mathcal{M}_{\mathcal{P}_1}^{(i_2)}+\sum_{j=2}^{4}\mathcal{M}_{\mathcal{P}_j}^{(i_1)}$. From these two quantities, $\mathcal{A}$ can obtain  $\mathcal{M}_{\mathcal{P}_1}^{(i_2)} - \mathcal{M}_{\mathcal{P}_1}^{(i_1)}$, which violates the privacy of the local model for $\mathcal{P}_1$. In \textit{DeTrust-FL}, encrypted model updates
are labeled with a round ensuring that the aggregator needs to use decryption keys only for the current round. Therefore, the aggregator cannot use ciphertexts from previous rounds in any round, which prohibits inference via replay attacks.

Based on the threat model described in Section~\ref{sec:ht}, we do not consider the DDoS attacks or malicious parties. 
Thus, if the \textit{Party-Inspect-Negotiate} verification is passed, the party does not intentionally refuse the training query.

\subsection{Forced Secure Aggregation}

Existing secure aggregation solutions that employ homomorphic encryption schemes (e.g., Paillier and threshold Paillier) cannot ensure that parties' encrypted models are included in the aggregated model.
In these solutions, the aggregator is responsible for the fusion of the encrypted model updates. Simultaneously, the parties perform the decryption or partial decryption such that a \textit{curious} aggregator can manipulate the fusion operation by only involving one party's encrypted model update.
This is a form of \textit{isolation attack} \cite{chen2011secure, liu2019boosting}.
While it may not result in privacy leakage for some secure aggregation schemes, e.g, one based on Paillier encryption, it may result in a biased global model or aggregated model that does not generalize well to the target party's data.

Unlike existing solutions, \textit{DeTrust-FL} enables the forced secure aggregation functionality, where each party can ensure that its encrypted local model update is included in the aggregated model.
\textit{DeTrust-FL} does not involve any ciphertext fusion operation. 
To acquire the aggregated model, the aggregator performs the decryption over a set of encrypted local models while the decryption key is associated with the fusion weight vector that all parties should agree on.
Thus, a party can check its assigned weight in the participation matrix.
Suppose that the aggregator manipulates the fusion weight $\pmb{V}_{i}$ for target party $\mathcal{P}_{j}$ by sending other parties $\pmb{V}^{\text{manipulate}}_{i}$ s.t. $v^{\text{manipulate}}_{i,j}=0$ while sending $\mathcal{P}_{j}$ the fusion weight $\pmb{V}^{\text{valid}}_{i}$ s.t. $v^{\text{valid}}_{i,j}=1$ to make $\mathcal{P}_{j}$ believe that its encrypted model is involved, but actually not. 
However, the function decryption key fragments $\mathtt{dk}^{\pmb{V}^{\text{valid}}_{i}}_{\mathcal{P}_{j}}$ generated by $\mathcal{P}_{j}$ and $\{\mathtt{dk}^{\pmb{V}^{\text{manipulate}}_{i}}_{\mathcal{P}_{k}}\}_{j\ne k}$ generated by others cannot collaboratively recover $\mathtt{dk}^{\pmb{V}^{\text{manipulate}}_{i}}$, which is ensured by the underlying cryptosystem.
As a result, such manipulation is impossible.

\section{Experimental Evaluation}
\label{sec:eval}

\subsection{Experimental Setup}

The implementation of the \textit{DeTrust-FL} framework builds on the IBM Federated Learning (IBM-FL) framework \cite{ludwig2020ibm}. 
We  implement the underlying cryptosystem, i.e., the decentralized MCFE scheme \cite{abdalla2019decentralizing} in Python based on the \textit{gmpy2} library.
We also incorporate the techniques for final discrete logarithm computation as used in \cite{xu2019hybridalpha} in our implementation. 
To integrate our proposed solution with the IBM-FL framework, we implement \textit{DeTrust-FL} privacy-preserving \textit{global fusion handler} and \textit{local training handler} classes for both aggregator and parties that respectively inherited the \textit{IterAvgFusionHandler} class and \textit{LocalTrainingHandler} class of IBM-FL.
We adopted and implemented an instance of DMCFE of \cite{abdalla2019decentralizing}.

\begin{table}[h]
    \centering
    \begin{threeparttable}
    \footnotesize
    \caption{Summary of Keras Model, Dataset Assignment and FL Setting}
    \label{table:setting}
    \begin{tabular}{lrcccc}
        \toprule
            Type$\dagger$ 
            & Parameters & Parties & Training/Test per Party & FL Rounds & Local Epochs \\
        \midrule
        CNN-MNIST 
        & 1,199,882 & 5 & 500/2000 (non-iid) & 20 & 3 \\
        \hline
        CNN-CIFAR10 
        & 890,410 & 10 & 5000/1000 (non-iid) & 30 & 20\\
        \bottomrule
    \end{tabular}
    \begin{tablenotes}
    \item[$\dagger$] We use different CNN architecture for different datasets. For MNIST, we use 2xConv2D$\to$MaxPooling$\to$Droupout$\to$Flatten$\to$Dense$\to$Dense while for CIFAR10 we use 2x(2xConv2D$\to$MaxPooling$\to$Droupout)$\to$Flatten$\to$Dense$\to$Dense.  
    \end{tablenotes}
    \end{threeparttable}
\end{table}

To benchmark the performance of \textit{DeTrust-FL}, we train two types of Keras-based Convolutional Neural Network (CNN) models, as described in \tablename~\ref{table:setting}, to classify the publicly available MNIST dataset of handwritten digits \cite{yann2010mnist} and CIFAR10 dataset of colour images \cite{krizhevsky2009learning}.
The total parameters of two CNN models are 1,199,882 and 890,410, respectively.
\tablename~\ref{table:setting} also shows dataset assignment for each party in non-iid setting.

In the configuration of the IBM-FL framework, as partially included in \tablename~\ref{table:setting}, we set the following hyperparameters: 
(i) termination accuracy is 0.99; 
(ii) total global training rounds is 20 for MNIST dataset and 30 for CIFAR10 dateset; 
(iii) learning rate is 0.01;
(iv) each local training epoch varies according to the evaluated dataset and adopted model.
The crypto-based secure aggregation works on the \textit{integer} field, while model parameters are in \textit{floating-point number} format.
\textit{DeTrust-FL} has a parameter of \textit{encoding precision} to define the scale factor of the conversion between integers and floating-point numbers, as do in existing solutions \cite{xu2019hybridalpha, zhang2020batchcrypt, liu2019secure}.
We use the term precision to denote the number of digits after the decimal point that are considered for rounding off. The default encoding precision is 4, which means we keep four digits after the decimal point.  The rounded off number is then appropriately scaled to convert it to an integer.

As analysed in Section~\ref{sec:comparison-with-existing}, the secure aggregation is easily integrated with differential privacy (DP) technique in the hybrid methodology.
Because the impact and effect of DP-integration are reported in \cite{xu2019hybridalpha,truex2019hybrid}, we omit the similar results
for the sake of brevity.
We compare our \textit{DeTrust-FL} framework to various baselines:
\begin{itemize}
    \item General FL training without any secure aggregation setting (General-FL);
    \item FL training using partially additive homomorphic encryption (i.e., Paillier cryptosystem) based secure aggregation (PHE-FL, e.g., \cite{liu2019secure, sharma2019secure});
    \item FL training using threshold Paillier based secure aggregation (HybridOne \cite{truex2019hybrid});
    \item FL training using functional encryption based secure aggregation (HybridAlpha  \cite{xu2019hybridalpha}). 
\end{itemize}
Note that, while both \textit{DeTrust-FL} and \textit{HybridAlpha} employ functional encryption techniques,
\textit{HybridAlpha} relies on a trusted online third-party entity to provide key service for each training round.

All experiments are performed on a cloud instance equipped with Intel(R) Xeon(R) CPU E5-2683 v4 platform with 32 cores and 64GB of RAM.
Note that the network latency is not measured in our experiment as the framework is running on the same machine in a multi-process setting.

\subsection{Experimental Results}

\begin{figure*}[htb]
    \centering
    \includegraphics[width=\textwidth,trim=16 12 16 10, clip]{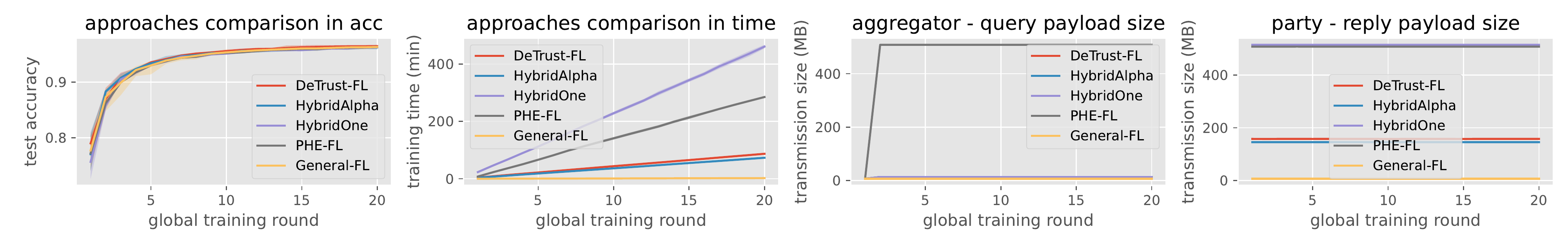}
    \caption{Model accuracy, training time and transmission payload comparison for multiple secure aggregation schemes in FL training on evaluating MNIST dataset.}
    \label{fig:cmp_diff_approaches}
\end{figure*}

\subsubsection{Comparison of Performance With Various Approaches}


\noindent\textbf{Model Accuracy}.
We present the model accuracy of \textit{DeTrust-FL} with comparison to four baselines using the MNIST dataset. 
Each test case is repeated at least three times, and we report the min/max/avg values in our results.
As shown in \figurename~\ref{fig:cmp_diff_approaches}, we report the model accuracy of the global model over 20 global training rounds, where for each training round each party  performs 3 local training epochs.
We observe that  
(i) the introduction of secure aggregation does not impact the final model accuracy and the rate of learning convergence, and
(ii) \textit{DeTrust-FL} has comparable model quality with respect to the general FL training without secure aggregation setting, as well as to compared FL frameworks with secure aggregation.

\noindent\textbf{Training Time}.
We also analyze the impact of the secure aggregation approaches on total training time, as shown in \figurename~\ref{fig:cmp_diff_approaches}, the second figure.
It shows that \textit{DeTrust-FL} has a training time close to \textit{General-FL}  and achieves significant training efficiency improvements compared to existing secure aggregation-enabled FL solutions.

\textit{DeTrust-FL} achieves  efficiency comparable to  \textit{HybridAlpha}, with both being the most efficient of the crypto-based secure aggregation approaches. 
However, \textit{HybridAlpha} relies on a centralized and trusted third-party authority to do so while  \textit{DeTrust-FL} does not.
In the context of 20 global training rounds, \textit{DeTrust-FL} reduces training time by about 70.2\% and 80.3\% when compared to secure aggregation solutions of \textit{PHE-FL} and \textit{HybirdOne} frameworks, respectively.
The underlying cryptosystem and reduced communication interactions contribute to this improvement in training time efficiency, as the \textit{DeTrust-FL} utilizes decentralized multi-input client encryption with more efficient encryption and decryption algorithms, and only requires one round of interaction per secure aggregation.

\noindent\textbf{Communication Interaction and Transmission Overhead}.
To measure communication efficiency, first we theoretically analyze the communication rounds needed by each approach and then report the  experimental results.

Suppose that the FL training is performed over $m$ 
rounds of global training among $n$
parties $\mathcal{P}$, a key server $\mathcal{K}$ and a aggregator $\mathcal{A}$.
Notice that, in \textit{DeTrust-FL}, the key server participates only during the setup time.
Table~\ref{table:cmp_comm} summarizes total number of interactions. 
It shows that \textit{DeTrust-FL} uses the smallest number of interactions compared to other secure aggregation solutions.
Note that we include one round of registration interaction between the aggregator and $n$ parties, as required in the IBM-FL framework.

We also accumulate the total transmission payload for each FL training round, in terms of aggregator's query and parties' response, with respect to all approaches.
\figurename~\ref{fig:cmp_diff_approaches} reports the comparison of transmission payload during the entire FL training. 
Compared to \textit{PHE-FL} and \textit{HybridOne} solutions, \textit{DeTrust-FL} reduces the volume of transmission payload by 73.6\% and 82.2\%, respectively.
Our work also generates a  transmission payload similar to \textit{HybridAlpha}. However, we reduce the number of interactions by 16.4\% in the setting of 20 global training rounds with 5 parties, as shown in Table~\ref{table:cmp_comm}.

\begin{table}[htb]
    \scriptsize
    \centering
    \begin{threeparttable}
    \caption{Comparison of communication interaction}
    \label{table:cmp_comm}
    \begin{tabular}{lcccc}
        \toprule
            Proposal & $\mathcal{A} \leftrightarrow \mathcal{P}$ & $\mathcal{A} \leftrightarrow \mathcal{K}$ & $\mathcal{P} \leftrightarrow \mathcal{K}$ & Total \\
        \midrule
            General-FL & $mn+n$ & 0 & 0 & $mn + n$\\
            PHE-FL & $2mn+n$ & 1 & $n$ & $2mn+2n+1$\\
            HybridOne & $2mn+n$ & 1 & $n$ & $2mn+2n+1$\\
            HybridAlpha & $mn+n$ & $m+1$ & $n$ & $mn+m+2n+1$\\
            \textbf{DeTrust-FL}  & $mn+n$ & 1 & $n$ & $mn+2n+1$\\
        \bottomrule
    \end{tabular}
    \begin{tablenotes}
      \small
      \item Notation: $m$ rounds of global training with one aggregator $\mathcal{A}$, $n$ parties $\mathcal{P}$ and one key server (or TPA) $\mathcal{K}$.
    \end{tablenotes}
    \end{threeparttable}
\end{table}

\subsubsection{Performance of DeTrust-FL}
As demonstrated above, \textit{DeTrust-FL} and \textit{HybridAlpha} solutions outperform all other solutions in terms of accuracy and training time.
We now further examine the performance of \textit{DeTrust-FL} in the context of 30 FL training rounds with 10 parties using the CIFAR10 dataset.

\begin{figure*}[htb]
    \centering
    \centering
    \includegraphics[width=\textwidth,trim=12 12 15 10, clip]{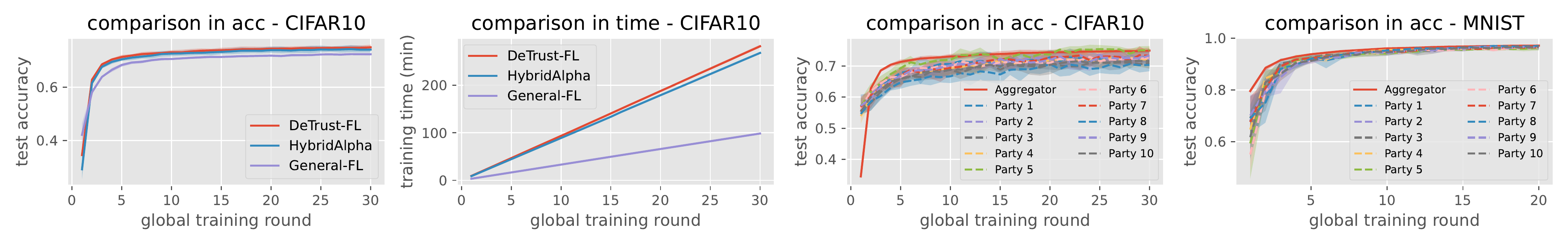}
    \caption{Performance comparison on evaluating CIFAR10 dataset and decomposition of model accuracy in \textit{DeTrust-FL} training on evaluating CIFAR10 dataset and MNIST dataset.}
    \label{fig:performance_detrust}
    \vspace{-3mm}
\end{figure*}

We first report the model performance of \textit{DeTrust FL}, as well as \textit{General-FL} and \textit{HybridAlpha}. We evaluate model accuracy and training time in the setting as illustrated in \tablename~\ref{table:setting} with encoding precision as four on floating-point parameters. 
As shown in the first two figures of \figurename~\ref{fig:performance_detrust}, we report the model accuracy curve and training time for CIFAR10 dataset.
We observe that both \textit{DeTrust-FL} and \textit{HybridAlpha} can achieve better model accuracy compared to the \textit{General-FL}.
Theoretically, we believe that the slight improvement in accuracy is due to the encoding operation that converts the model weights between integers and floating-point numbers, which discards some information and could be considered as a type of pruning.
However, we left it as an open question to be investigated further in the future.

As shown in last two \figurename~\ref{fig:performance_detrust}, we also illustrate that \textit{DeTrust-FL} inherits the advantage of general FL wherein the global model has a better rate of convergence and model accuracy compared to those local models, for both MNIST and CIFAR10 dataset.

\subsubsection{Impact of Numbers of Parties}
\label{sec:res:parties}

\begin{figure}[t]
    \centering
    \includegraphics[width=0.6\textwidth,trim=16 12 16 10, clip]{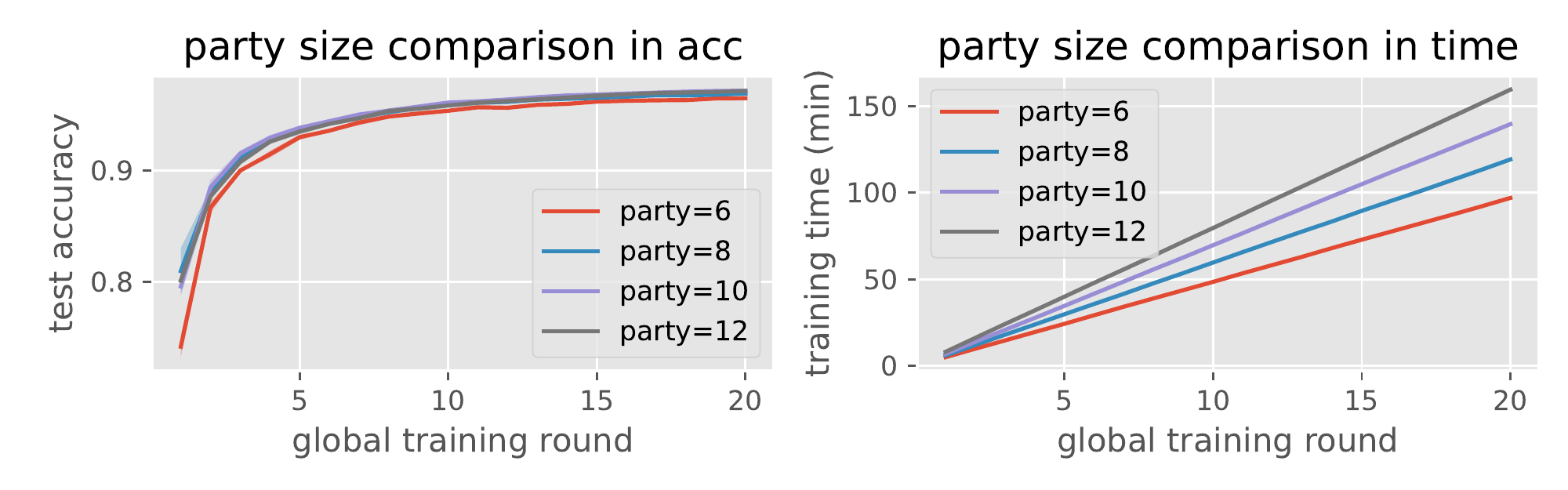}
    \caption{Impact of number of parties in \textit{DeTrust-FL} training on evaluating MNIST dataset with settings of precision $=4$ and 3 local training epochs per training round.}
    \label{fig:cmp_parties}
\end{figure}

We also evaluate the impact of numbers of parties on \textit{DeTrust-FL} performance.
As reported in \figurename~\ref{fig:cmp_parties}, more parties enrolled in the FL training can increase the model accuracy. 
In our experimental setting, each party's training samples are assigned randomly; thus, the model accuracy improvement is not as significant as if they were more unbalanced, which often occurs in practice.
\figurename~\ref{fig:cmp_parties} also shows the training time results over 20 global training rounds. 
The training time increases along  the number of parties.
Each party in the IBM FL experiment is simulated by multiple processes in a parallel setting on a single node. Hence, the increased training time is caused by the increased computation time in the decryption phase and functional decryption key recovering phase at the aggregator, rather than the time spent at the parties.

\subsubsection{Impact of Encoding Precision on Floating-point Parameters}
\label{sec:res:precision}

\begin{figure}[t]
    \centering
    \includegraphics[width=0.8\textwidth,trim=16 12 16 10, clip]{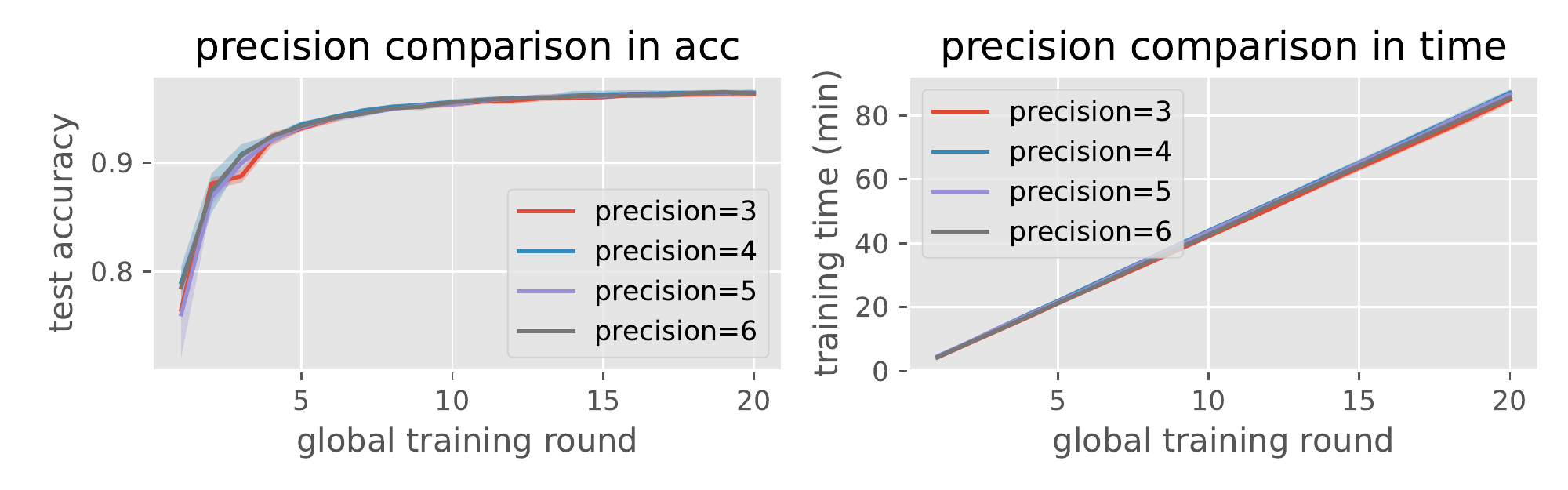}
    \caption{Impact of encoding precision of floating-point parameters in \textit{DeTrust-FL} training on evaluating MNIST dataset with setting of 5 parties and 3 local training epochs per training round.}
    \label{fig:cmp_precision}
\end{figure}

When performing crypto-based secure aggregation, the floating-point model parameters should be encoded in integer format. We evaluate the impact of encoding precision on model accuracy and training time in \textit{DeTrust}. 
As shown in \figurename~\ref{fig:cmp_precision}, the encoding precision setting has no significant influence on model accuracy.
Furthermore, a higher encoding precision setting may slightly increase the training time.
As mentioned above, since parties are simulated with multiple processes in this experiment there are some fluctuations in the growth curve.

\section{Related Work}
\label{sec:related}

Federated learning (FL) as proposed in \cite{konevcny2016federated} has emerged as a promising approach to collaboratively train a global model without sharing local data among participants and thereby providing a primary privacy guarantee.
The primary FL design is not sufficient for privacy-sensitive applications because it is still vulnerable from inference attacks on the final model and model updates between parties and aggregator, as  demonstrated by \cite{nasr2019comprehensive, shokri2017membership}.
To partially address those threats, several privacy-preserving techniques such as differential privacy and secure multi-party computation have been adopted during the learning process.

Emerging hybrid privacy-preserving methodologies have shown promise to providing strong privacy guarantees \cite{xu2019hybridalpha, truex2019hybrid}. The approach in \cite{truex2019hybrid} employs the threshold Paillier cryptosystem to perform multi-party secure aggregation, while \textit{HybridAlpha} \cite{xu2019hybridalpha} uses a functional encryption cryptosystem to securely aggregate encrypted local models.
Existing secure aggregation approaches mainly follow three directions: \textit{pairwise masking} based approach \cite{bonawitz2017practical,kadhe2020fastsecagg}, \textit{homomorphic encryption} based approach \cite{ryffel2018generic,truex2019hybrid,liu2019boosting} and \textit{functional encryption} \cite{xu2019hybridalpha,xu2020revisiting,xu2019cryptonn,xu2021nn,xu2021fedv}.

Pairwise masking \cite{bonawitz2017practical,kadhe2020fastsecagg} is a lightweight computation approach where each input is masked with one-time pads while the aggregated output cancels out those pads. 
For example, Bonawitz et al. \cite{bonawitz2017practical} make use of double pairwise additive masking by integrating approaches such as secret sharing, key agreement, authenticated encryption, and public key infrastructure.
Even though those approaches are efficient in computation, they either require peer-to-peer communication or rely on a server forwarding the encrypted shares and masked inputs for each aggregation round, resulting in communication inefficiency.
Additionally, they are vulnerable to disaggregation attacks that are recently demonstrated in \cite{lam2021gradient, so2021securing} in the context of multi-round secure aggregation in FL training.

In homomorphic encryption-based approaches, one set of solutions such as \cite{truex2019hybrid,liu2019boosting} rely on additive homomorphic encryption schemes like the Paillier and its variants.
Another set of solutions such as PySyft \cite{ryffel2018generic} employs the SPDZ protocol that is built on somewhat homomorphic encryption (SHE) in the form of BGV \cite{brakerski2014leveled}.
Those approaches are efficient in communication but are computationally  inefficient. 
Both pairwise masking and homomorphic encryption-based secure aggregation solutions only support average fusion algorithms and are not applicable to weighted fusion algorithms.
In addition, homomorphic encryption-based secure aggregation cannot guarantee secure aggregation to a party without special hardware. 

A functional encryption-based solution, \textit{HybridAlpha}, \cite{xu2019hybridalpha} has achieved significant training efficiency improvements by efficient cryptographic computation and simple communication design.
Furthermore, it also supports the weighted secure aggregation situation and addresses
ensures a minimum number of pre-defined replies are aggregated.
However, \textit{HybridAlpha} introduces and requires an online third-party entity that should be fully trusted by all parties, which makes impractical in some FL scenarios.

None of the approaches outlined above achieves computational and communication efficiency without requiring a trusted crypto authority to provide cryptographic key services or a partially trusted aggregator to honestly execute secure aggregation methods and without conducting stealth targeted or disaggregation attacks.

\section{Conclusion}
\label{sec:conclusion}
Secure aggregation, the underlying privacy-preserving computation primitive, plays an essential role during the privacy-preserving FL training. 
Existing approaches do not address the dilemma of efficiency requirements and decentralized trust setting.
To address these limitations, in this paper
we propose \textit{DeTrust-FL}, a novel FL framework that does not require a fully trusted crypto dealer, as well as mitigating partial trust on the aggregator.
In addition, our experimental evaluation shows that our approach achieves comparable performance in terms of model accuracy and training time, and reduces the communication interactions by 16.4\% compared to the state-of-the-art efficient approach without the prerequisite of a centralized trust setting.

\bibliographystyle{plain}
\bibliography{ref}

\end{document}